\documentstyle[11pt]{article}
\input psfig
\long\def\comment#1{}
\begin{document}
\title{Information Complexity of Quantum Gates}
\author{Subhash Kak\\
Department of Electrical \& Computer Engineering\\
Louisiana State University,
Baton Rouge, LA 70803, USA}
\maketitle

\begin{abstract}

This paper considers the realizability of quantum gates from the
perspective of information complexity.
Since the gate is a physical device
that must be controlled classically, it is subject to random error.
We define
the complexity of gate operation in terms of the difference between
the entropy of
the variables associated with initial and final states of the computation.
We argue that the gate
operations are irreversible if there is a difference in the 
accuracy associated with input
and output variables.
It is shown that under some conditions
the gate operation may be associated with unbounded entropy, implying
impossibility of implementation.

\end{abstract}

\thispagestyle{empty}

\subsection*{Introduction}

In this paper, we consider complexity and realizability
of quantum gates from the point of view of
information theory. 
A gate is a physical system that is controlled by varying some input variables,
which
are classical.
In principle, such a physical system could implement a variety of operators based on
the control variables.
The gate functions may be also implemented by 
a single physical system that
operates sequentially on the qubits in the quantum register.
The complexity of the gate will be defined in terms of the
entropy associated with its control.
From a practical point of
view, one is interested in asking how
easy it is to control a gate.

As no analog system can have infinite precision, we investigate
what happens if the precision levels at the input and the output are different.
The complexity of the gate, defined in terms of entropy, will
be examined
for the rotation and {\sc cnot} gates in certain circuits.

\subsection*{Information processing by gate}
One aspect of gate performance is its accuracy.
Researchers on quantum information science have given much attention to
the question of errors and their correction [1-3] by drawing upon parallels with
classical information.
Quantum
error-correction coding works like classical 
error-correction to correct some large errors.

But the framework
of quantum information is distinct from that
of classical information.
In the classical case, 
it is implicitly assumed that there occurs an
automatic correction of errors that are
smaller than
a threshold by means of clipping 
or by the use of a decision circuit.
In the case of quantum
information, the input data is nominally
discrete, but in reality its precision cannot be absolute in any actual
realization.
Furthermore, unknown small errors
in quantum information
cannot be corrected [4-5].
Consequently, proposals for error correction  and fault tolerance (such as [6-8])
remain unrealistic.

Classical analog computation and quantum processing
do have parallels. In
general, fixed errors in gate operation could become irreversible due
to actual small nonlinearity of nominally linear elements.
Analog computing is not practical to implement because
noise cannot be separated from useful signal and it accumulates, degrading
the system performance in an uncorrectable manner.

If there were no noise, the practicality of analog computing would
depend on the feasibility of the gate implementation over
the expected input-output range.
This feasibility
must be checked in the context of the limitations on information processing
by the gate.

Consider the gate G of Figure 1. 
It may be assumed that it is a physical system which is controlled by means
of some variable. This control is implemented
by choosing a setting on an instrument, and this
choice is associated
with random error. If one views the circuit operations to be implemented by
the same device transitioning through various states in sequence, then one can
determine the distribution of the control variable states, and compute its
entropy. This entropy, when determined for the entire computing circuit, may
be taken to represent its complexity.

\begin{figure}
\hspace*{0.2in}\centering{
\psfig{file=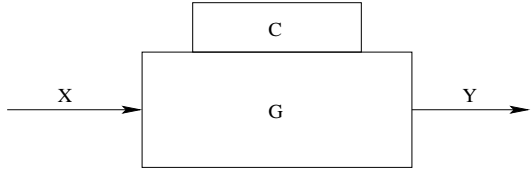,width=7cm}}
\caption{Information processing gate, $G$, with control, $C$}
\end{figure}
Information is preserved, therefore
one can define the following relationship
for the entropy expressions for the input $X$, the gate control information $C$,
and the output $Y$:

\begin{equation}
H(Y) = H(X) + H(C).
\end{equation} 

Although it is assumed that the variable $X$ is discrete, in reality the
lack of perfect precision at the state preparation state makes it a continuous
variable [9-11].
The lack of precision may not
affect the measurement variables, but it would introduce continuous phase error.

Similarly, the output variable $Y$ has discrete measurement 
associated with it, but it may come with additional component states and many
unknown, continuous phase terms. This has implications for quantum 
amplitudes and, consequently, with the
probabilities associated with the states.

As an aside, equation (1) provides an explanation for the no-cloning
theorem. A gate cannot clone a state since this would require the gate to 
supply information equal to that of the unknown state, which, by virtue of its
being unknown, is impossible.

As the variables $X$ and $Y$ (defined 
together with associated continuous phase terms)
are continuous, the classical
variable $C$ must also 
be continuous. The entropy associated with a continuous variable $Z$
is given by the expression:

\begin{equation}
H (Z) = h(Z) - \lim_{\Delta z \rightarrow 0} log_2 \Delta z 
\end{equation}

\noindent
where $h(Z)$ is the differential entropy:

\begin{equation}
h(Z) = \int_{- \infty}^{\infty} f_Z (z) log_2 \left[\frac{1}{f_Z (z)}\right] dz
\end{equation}

\noindent
and $\Delta z$ is the precision associated with the variable.

If the
precision is the same at both input and output,
the term $\lim_{\Delta z \rightarrow 0} log_2 \Delta z$ will cancel out and the
differential entropies would be a proper measure of the entropy of $X$ and $Y$.
In other words,

\begin{equation}
H(C) = h(Y) - h(X).
\end{equation}

The entropy associated with H(C) is the information lost in the computation process and
it may be converted to heat according to thermodynamic laws [12-14].
If $H(C)$ is non-zero, error-free quantum computation is impossible, since
this
is associated with loss of information.

\subsection*{Multiplication by constant}
\noindent
{\bf Example 1.}
Consider a gate which multiples the inputs by a fixed constant $k > 1$. 
If the input $X$ is distributed uniformly over the interval $(0,a)$, then the
output $Y$ is distributed uniformly over $(0,ka)$. The differential entropy values 
of the input and the output are:

\begin{equation}
h(X) = log_2 a
\end{equation}

\begin{equation}
h(Y) = log_2 ka
\end{equation}

Assuming the same
precision at input
and output, the gate needs to supply 
entropy equal to $ H(C) = log_2 ka -log_2 a = log_2 k$, 
which
become large as $k$ increases.
This supply of entropy will have to be done in terms of interpolation or other
processing which cannot be perfect.

If $k < 1$, then the output entropy is smaller than input entropy and, therefore,
$H(C)$ represents loss of information in the output.
In effect, the assumption of fixed amplification of a variable with the same
absolute precision at the output amounts to a nonlinear, irreversible process.
For example, when a picture is compressed, one cannot obtain the
original to the earlier precision by amplifying it back.
In practical terms, the precision needed for the realization of a universal gate will
be unattainable for a variety of reasons: one cannot have perfectly
linear behavior in an electrical circuit over an unrestricted range.
Unrestricted multiplication of a continuous variable is not
implementable if the precision remains unchanged.

In quantum computing, problems that somewhat parallel this above example are
the implementation of rotation and 
{\sc cnot} gates, two operators that are basic to the computation process [15].
The necessarily classical control of the gate 
is marred by random errors as well
as
calibration errors. 

\subsection*{Rotation}

\noindent
{\bf Example 2.}
Consider a quantum gate that rotates the input qubit by a fixed angle.
Since the input $X$ and the output $Y$ will
be distributed uniformly over the same interval $(0,a)$, the
entropy associated with this gate will be $0$ (as per equation 4) 
as is required by the reversible nature of the assumed quantum
evolution.

But if the precision associated
with the measurement and initialization processes at the input and the output
is different, then lossless (or, equivalently, error-free) evolution cannot
be assumed.

\subsection*{{\sc cnot} and Hadamard gates}

Consider the {\sc cnot} gate together with a companion Hadamard
gate. The errors in the device implementation of the {\sc cnot}
gate may make
the gate effectively nonlinear and hence nonunitary. 
The matrix values that the device embodies may be different from the
nominal ones below:

\vspace{0.2in}
\begin{equation}
 \left[ \begin{array}{cccc}
1 & 0 & 0 & 0 \\
0 & 1 & 0 & 0 \\
0 & 0 & 0 & 1 \\
0 & 0 & 1 & 0 \\
\end{array} \right]
\end{equation}

\vspace{0.1in}

For simplicity, we consider a very straightforward situation which does not
affect the {\sc cnot} gate, but where
its companion
Hadamard gate is off the correct value,  stuck in the state
\vspace{0.2in}
\begin{equation}
                     H_s =   \left[ \begin{array}{cc}
                                  cos \theta  & sin \theta \\
                                  sin \theta & -cos \theta \\
                               \end{array} \right]
\end{equation}
where $\theta \neq 45^o$.

\subsubsection*{Stuck Hadamard gate before a {\sc cnot}}

\noindent
{\bf Example 3.}
Consider the arrangement of Figure 2, where the stuck gate $H_s$ 
($\theta \neq \pi /4$, but its value is known) is to
the left of the {\sc cnot} gate;
this circuit demonstrates that quantum processing
can compute a global property of a function by a single measurement [1].

\begin{figure}
\hspace*{0.2in}\centering{
\psfig{file=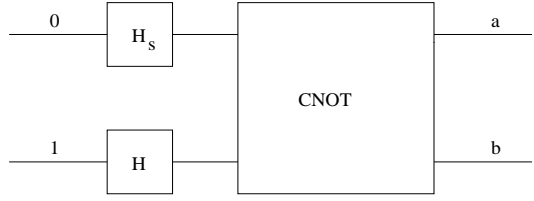,width=7cm}}
\caption{The stuck Hadamard gate $H_s$ before {\sc cnot}}
\end{figure}

\vspace{0.1in}

It will be seen that at the output of the {\sc cnot} gate, the
state is: 

\vspace{0.1in}

$ \frac{1}{\sqrt 2}~(cos \theta |0\rangle - sin \theta |1\rangle ) ~ (|0\rangle - |1\rangle)$

\vspace{0.1in}

The state $|a\rangle = (cos \theta |0\rangle - sin \theta |1\rangle )$,
which is in error, may be passed
through the gate
$                       \left[ \begin{array}{cc}
1 & 0  \\
0 & -1  \\
\end{array} \right]$ followed by
  another $H_s$ to yield $|0\rangle$, which can be transformed to the correct
$|a\rangle= \frac{1}{\sqrt 2} (|0\rangle - |1\rangle)$. In this example, the
state
$|b\rangle= \frac{1}{\sqrt 2} (|0\rangle - |1\rangle)$ was not affected by
the stuck gate $H_s$.

When the stuck gate is the lower Hadamard gate, as in Figure 3,
the state at the output of the {\sc cnot} gate is:

\vspace{0.1in}
\begin{figure}
\hspace*{0.2in}\centering{
\psfig{file=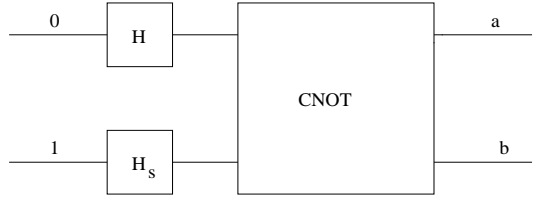,width=7cm}}
\caption{The stuck Hadamard gate $H_s$ before {\sc cnot} in the lower input}
\end{figure}

\vspace{0.1in}

$\frac{1}{\sqrt 2} (sin \theta |00\rangle - cos \theta |01\rangle 
+ sin \theta |11\rangle - cos \theta |10\rangle )$

\vspace{0.2in}
\noindent
Corresponding to this we have the density function $\rho^{ab}$ given below:

\vspace{0.2in}
$\rho^{ab} = \frac{1}{2}
        \left[ \begin{array}{cccc}
sin^2 \theta & -sin \theta cos \theta & -sin \theta cos \theta & sin^2 \theta \\
-sin \theta cos \theta & cos^2 \theta & \cos^2 \theta & -sin \theta cos \theta \\
-sin \theta cos \theta & \cos^2 \theta  & \cos^2 \theta  & -sin \theta cos \theta \\
sin^2 \theta & -sin \theta cos \theta & -sin \theta cos \theta & sin^2 \theta \\
\end{array} \right]$

\vspace{0.2in}

\noindent
It follows that the reduced density matrix for the state $|a\rangle$ is:

\vspace{0.2in}
 $  \rho^a =\frac{1}{2}                    \left[ \begin{array}{cc}
1 & -sin 2\theta  \\
-sin 2\theta & 1  \\
\end{array} \right]$

\vspace{0.2in}
Therefore, when $\theta \neq \pi /4 $, $\rho^{ab}$ is a mixture, and we cannot perform 
any local correction to $|a \rangle$ to obtain
the correct product state, for a unitary transformation on a mixture will keep it
as a mixture.
In other words, this error is not locally correctable.

\subsection*{Stuck Hadamard gate in the teleportation protocol}

\noindent
{\bf Example 4.}
In the teleportation protocol,
an unknown quantum state  (of particle $X$) is teleported to a remote location
using
two entangled particles  ($Y$ and $Z$) and 
classical information. Here, for convenience, we use the variant teleportation protocol 
[16] which
requires only one classical bit in its classical information link (Figure 4).
But instead of the Hadamard operator, we consider 
$H_s$ to be the rotation operator with angle $\theta$. 
We assume that the receiver has a copy of $H_s$ available for local 
processing, and we would like to estimate what would happen if this copy is
not identical to the one used at the transmitting end.

\vspace{0.1in}
\begin{figure}
\hspace*{0.1in}\centering{
\psfig{file=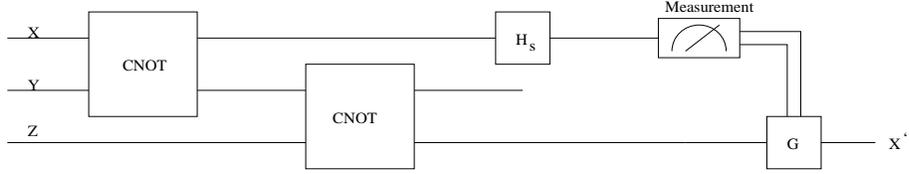,width=12cm}}
\caption{The stuck Hadamard gate $H_s$ in the teleportation set-up}
\end{figure}
\vspace{0.1in}

The state $X$ is $|\phi\rangle =  \alpha |0\rangle + \beta |1\rangle $, where 
$\alpha$ and $\beta$ are unknown amplitudes, and $Y$ and $Z$ are 
in the pure entangled state
$ \frac{1}{\sqrt 2} (| 00 \rangle~ +~  | 11\rangle )$.
The initial state of the three
particles is:

\vspace{0.2in}

$\frac{1}{\sqrt 2} (\alpha~  |000\rangle + \beta~ |100\rangle + 
\alpha~ |011\rangle + \beta~ |111\rangle $

\vspace{0.2in}

\noindent
The sequence of steps 
in Figure 4 is as follows:

\begin{enumerate}

\item Apply chained transformations: {\sc cnot} on $X$ and $Y$, followed by
{\sc cnot} on $Y$ and $Z$.

\item Apply $H_s$ on the state of $X$.

\item
Measure the state of $X$ and transfer information regarding it.

\item
Apply appropriate operator $G$ to complete teleportation of the
unknown state.

\end{enumerate}

\noindent
A simple calculation will show that the state before the measurement is:

\vspace{0.2in}
$  \frac{1}{\sqrt 2} |0\rangle (  |0\rangle +  |1\rangle )
 (\alpha~cos \theta~ |0\rangle~ +~ \beta~sin \theta~ |1\rangle )  +\frac{1}{\sqrt 2} |1\rangle ( |0\rangle~ +~  |1\rangle )
(\alpha~sin \theta~ |0\rangle~ -~ \beta~cos \theta~ |1\rangle )$

\vspace{0.2in}

\noindent
Therefore, after the measurement, we get either

\vspace{0.2in}
$ X^+ = \alpha~cos \theta~ |0\rangle + \beta~sin \theta~ |1\rangle$ 

\noindent
or

$X^- = \alpha~sin \theta~ |0\rangle - \beta~cos \theta~ |1\rangle$

\vspace{0.2in}
\noindent
based on whether the measurement was $0$ or $1$. Assuming
that the value of $\theta$ is also communicated to  it,   the receiver
can recover
 the unknown $X$ probabilistically;
when the value of $\theta$ is $45^o$, then the inversion is
trivially simple.

For simplicity, assume that the receiver needs to invert $X^+ $.
He will replicate Figure 4 at his end which means that the
Hadamard gate that he would use would have identical characteristics
(the same precision) to
the one used during the earlier operation.
He would now obtain either 

\vspace{0.1in}

$ X^{++} = \alpha~cos^2 \theta~ |0\rangle + \beta~sin^2 \theta~ |1\rangle$

\vspace{0.1in}
\noindent
or

\vspace{0.1in}
$ X^{+-} = \alpha~ |0\rangle + \beta~ |1\rangle$

\vspace{0.1in}
Similarly, $X^-$ will, in the next iteration, lead to:

\vspace{0.1in}
$ X^{-+} = \alpha~ |0\rangle + \beta~ |1\rangle$

\noindent
or

\vspace{0.1in}
$ X^{--} = \alpha~sin^2 \theta~ |0\rangle + \beta~cos^2 \theta~ |1\rangle$

\vspace{0.1in}

This procedure may be extended, and 
the probability of recovering the unknown state $X$ can be shown to be given by
the tree diagram of Figure 5.

\begin{figure}
\hspace*{0.2in}\centering{
\psfig{file=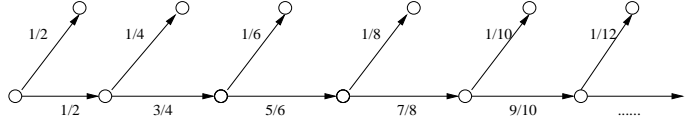,width=9cm}}
\caption{The probability tree for recovering the state $X$}
\end{figure}

In the first pass, there is a fifty percent probability of getting the correct state,
and this probability reduces in further passes (Figure 5).
The probability of recovering the state $X$ is thus:
\begin{equation}
\frac{1}{2} + \frac{1}{2}\frac{1}{4} + \frac{1}{2}\frac{3}{4}\frac{1}{6} +
\frac{1}{2}\frac{3}{4}\frac{5}{6} \frac{1}{8} +\frac{1}{2}\frac{3}{4}\frac{5}{6} \frac{7}{8} \frac{1}{10} + ...
\end{equation}

The ability of the receiver to implement the needed transformation will depend on the 
precision available in its gate control mechanism.
If the value of $\theta$ at
the sending point is smaller than the precision available
to the receiver, then the state $X$ cannot be recovered.

It is interesting that as long as the 
receiver possesses a rotation operator 
$H_s$ that is identical to
the one used at the sending point,
there is no need to 
know the value of $\theta$ and still obtain the unknown state $X$ probabilistically,
as in expression (9).

\subsection*{Conclusion}

We have considered the problem of gate complexity in quantum systems. 
The control of
the gate -- a physical device -- is by modifying some
classical variable, which is subject to error. Since one cannot assume
infinite precision in any control system, the implications of varying
accuracy 
emongst different gates becomes an
important problem.

We have
shown that in certain arrangements a stuck fault cannot be reversed down
the circuit stream using a single qubit
operator, for it converted a pure state into a mixed state.

We considered the case of the teleportation circuit 
with the rotation gate stuck at $\theta$.  When
$\theta = 0^o$, the state $X$ collapses to $0$ or $1$. When
$\theta \neq 0^o$ or $ 90^o$,
one may obtain the unknown state back probabilistically by passing $X^+$ or
$X^-$ back
through the circuit of Figure 4 iteratively.

Consider two parties, $A$ and $B$, who are both presented with the state $X^+$
or $X^-$.
If the precision available to one of them is greater than or equal to that of
the sender, and that of the other is less, then one of them can recover the
state, whereas the other cannot.

It is essential that the entropy rate associated with the quantum circuit
be smaller than what can be implemented by the information capacity of
the controller.
This perspective may be useful in evaluating proposals [17] for quantum computing
with noisy components.

\section*{References}
\begin{description}

\item
[1]
M.A. Nielsen and I.L. Chuang, {\it Quantum Computation and Quantum Information}.
Cambridge University Press, 2000.

\item
[2]
A.Y. Kitaev, ``Quantum computations: algorithms and error correction.''
{Russ. Math. Surv.}, 52, 1191-1249 (1997).

\item
[3]
E. Knill and R. Laflamme, ``A theory of quantum error-correcting codes.''
{\it Phys. Rev. A,} 55, 900-906 (1997).

\item
[4] S. Kak, ``General qubit errors cannot be corrected.''
{\it Information Sciences,} 152, 195-202 (2003); quant-ph/0206144.
\item
[5] S. Kak, ``The initialization problem in quantum computing.''
{\it Foundations of Physics,} 29, 267-279 (1999); 
quant-ph/9805002.

\item
[6]
A.M. Steane, ``Efficient fault-tolerant quantum computing.''
{\it Nature,} 399, 124-126 (1999).

\item
[7] E. Knill, ``Fault tolerant post-selected quantum computation."
Physics Arxiv: quant-ph/0404104.

\item
[8] K. Svore, B.M. Terhal, D. P. DiVincenzo, ``Local fault-tolerant
quantum computation.'' Physics Arxiv: quant-ph/0410047.

\item
[9]
S. Kak, ``Rotating a qubit.''
{\it Information Sciences,} 128, 149-154 (2000); quant-ph/9910107.

\item
[10]
S. Kak, ``Statistical constraints on state preparation for a quantum
computer.''
{\it Pramana,} 57, 683-688 (2001); quant-ph/0010109.

\item
[11]
S. Kak, ``Are quantum computing models realistic?''
Physics Arxiv: quant-ph/0110040.

\item
[12]
R. Landauer, ``Irreversibility and heat generation in the computing process.''
{\it IBM J. Res. Dev.,} 5, 183 (1961).

\item
[13]
C.H. Bennett, ``The thermodynamics of computation -- a review.''
{\it Int. J. Theor. Phys.,} 21, 905-940 (1982).
\item
[14]
S. Kak, ``Quantum information in a distributed apparatus.''
{\it Foundations of Physics} 28, 1005 (1998); Physics Archive: quant-ph/9804047.

(1998); quant-ph/9804047.
\item
[15]
D.P. DiVincenzo, ``Two-bit gates are universal for
quantum computation.'' 
{\it Phys. Rev. A,} 51, 1015-1022 (1995).

\item
[16] S. Kak, ``Teleportation protocols requiring only one classical bit."
Physics Arxiv: quant-ph/0305085.

\item
[17] E. Knill, ``Quantum computing with very noisy devices.''
Physics Arxiv: quant-ph/0410199.

\end{description}
 
\end{document}